# AUTOMATIC STRUCTURING OF SEMANTIC WEB SERVICES – AN APPROACH

B. Kamala
Asst Prof Gr – II, MCA Dept., Sri Sai Ram Engineering College, Chennai.
kamala.mca@sairam.edu.in

J. M. Nandhini
Asst Prof Gr – I, MCA Dept., Sri Sai Ram Engineering College, Chennai.
nandhini.mca@sairam.edu.in

## ABSTRACT

Ontologies have become the effective modeling for various applications and significantly in the semantic web. The difficulty of extracting information from the web, which was created mainly for visualising information, has driven the birth of the semantic web, which will contain much more resources than the web and will attach machine-readable semantic information to these resources. Ontological bootstrapping on a set of predefined sources, such as web services, must address the problem of multiple, largely unrelated concepts. The web services consist of basically two components, Web Services Description Language (WSDL) descriptors and free text descriptors.

The WSDL descriptor is evaluated using two methods, namely Term Frequency/Inverse Document Frequency (TF/IDF) and web context generation. The proposed bootstrapping ontological process integrates TF/IDF and web context generation and applies validation using the free text descriptor service, so that, it offers more accurate definition of ontologies. This paper uses ranking adaption model which predicts the rank for a collection of web service documents which leads to the automatic construction, enrichment and adaptation of ontologies.

## 1. INTRODUCTION

Ontology is the task of building computable models of some domain for some purpose and it is a specification of conceptualization. To be successfully used within information systems, ontologies should provide clear semantics for their concepts and a rich formalization of their semantics. These requirements are particularly important in the area of semantic web services whose ontological descriptions should be used for reasoning tasks. However, few existing ontologies fulfill these requirements.

The web is no longer just a set of documents; services now feature prominently, and both research and industry devote considerable effort and interest to service oriented architecture (SOA), web services, and related technologies. In a web of services or a service-oriented architecture, discovery is an essential task for building and using applications. Unfortunately the widely used service description techniques only cover the syntactic level.

Generic web service ontology should be used to describe any web services, independently of their domain. Some of the advantages of ontologies are,

- Allows to access and share more data.
- Improves accuracy and promotes completeness
- More flexible than simple tags
- Provide semantics to the information "between the tags"

A major requirement for web service ontology is high quality. In particular, the ontology should provide a high modeling expressiveness so that a large variety of services can be modeled. Clear semantics and a rich formalization of these semantics would ensure the support for complex reasoning tasks performed by services. Then, another measure of quality is whether the captured generic knowledge is adaptable for use in similar domains. It would also be desirable that a mapping between different evolving generic ontologies can be achieved in order to ensure a basis for their comparison.

Ontologies for web services should provide broad domain coverage and they should contain the knowledge that describes large and dynamic web service collections. [3]

The main existing solution for service discovery is UDDI (Universal Description, Discovery and Integration1), a standard for programmatically publishing and retrieving structured information about web services. Some companies operate public UDDI repositories, but its success has not been widespread because of complicated registration, missing monitoring facilities, and other difficulties. A few portals have been designed to act as web service repositories, but they all rely on manual registration and review, so coverage is limited and information falls out of date easily. Furthermore, the standard web search engines do not provide effective ways to search for services or to allow filtering according to availability and service parameters.

Figure 1.1 illustrates how a customer can access a service using UDDI registry.

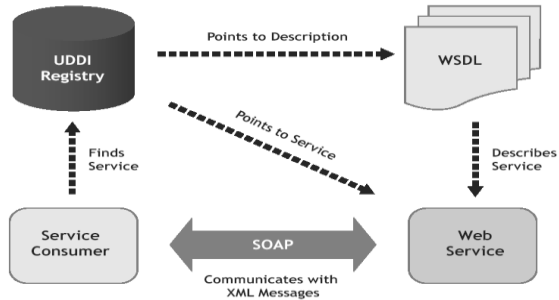

Figure 1.1 Web Service Access using UDDI

The remainder of this paper is organized as follows: Section 2 deals with the role of ontologies in web services. Section 3 discusses the approach of ontology bootstrapping. Section 4 discusses the role of context analysis in ontological bootstrapping. Section 5 deals with the need for generating onlogies and predicting rank for web services.

## 2. ROLE OF ONTOLOGIES IN WEB SERVICES

A key issue in enabling automatic interoperation among web services is to semantically mark up the services with shared ontologies. These ontologies typically fall into two categories: service ontology and domain ontology. The service ontology provides generic framework and language constructs for describing the modeling aspects of Web services, including process management, complex service composition, and security enforcement. The domain ontology describes concepts and concept relationships in the application domain, and facilitates the semantic markups on the domain-specific aspects of web services such as service categories, semantic types of parameters, etc. Clearly, such semantic markups are crucial to the interoperation of the Web services. [4]

Ontologies, however, are logical systems that themselves incorporate semantics. Formal semantics of knowledge representation systems allow user to interpret ontology definitions as a set of logical axioms. The ontology itself resolves inconsistencies and they do not need to do anything about them in the evolution framework. [6]

Ontology is an explicit specification of a conceptualization of a domain. Therefore, changes to any of the three elements in the definition can cause changes in ontology:
      (1) Changes in the domain
      (2) Changes in conceptualization or
      (3) Changes in the explicit specification.

Changes in the domain are very common and it requires major modifications in their respective schemas. Changes in conceptualization can result from a changing view of the world and from a change in usage perspective. Changes in the explicit specification occur when ontology is translated from one knowledge-representation language to another. [5]

Semantic alignment between ontologies is a necessary precondition to establish interoperability between agents or services using different ontologies. Hence, there remains the need to automatically combine multiple diverse and complementary alignment strategies of all indicators, i.e. extensional and intensional descriptions, in order to produce comprehensive, effective and efficient alignment methods. Such methods need to be flexible to cope with different strategies for various application scenarios.

**2.1 Use of Ontologies on Web Services**
- Improves the accuracy of web searches by searching for concepts instead of keywords.
- Allows systems that were independently developed to work together to exchange information.
- Facilitates the use of services to collect, process, and exchange information.
- Helps tackle complicated questions whose answers do not reside on a single web service.

**2.2 Ontology Convention in Agriculture and Finance**

Precision agriculture requires the collection, storage, sharing and analysis of large quantities of spatially referenced data. For this data to be effectively used, it must be transferred between different hardware, software and organizations. These data flows currently present a hurdle to uptake of precision agriculture as the multitude of data models, formats, interfaces and reference systems in use result in incompatibilities. [7]

Web services for professional-grade financial market data and applications fulfilling their customers request effectively. Many web services offers the broadest selection of financial Web services available today with multiple solutions covering domestic and global equities, commodities, currencies, fixed income and interbank interest rates, etc. [8]

## 3. BOOTSTRAPPING ONTOLOGY

The ontology bootstrapping process is based on analyzing a web service using three different methods, where each method represents a different perspective of viewing the web service. The Term Frequency/Inverse Document Frequency (TF/IDF) method analyzes the web service from an internal point of view, i.e., what concept in the text describes the WSDL document content. In the web service ontology bootstrapping process a web service can be separated into two types of descriptions:
- The Web Service Description Language (WSDL) describing "how" the service should be used.
- A textual description of the web service in free text describing "what" the service does. [2]

The bootstrapping ontology model proposed in this paper is based on the continuous analysis of WSDL documents and employs an ontology model based on concepts and relationships. The innovation of the proposed bootstrapping model focuses on the combination of the use of two different extraction methods, TF/IDF and web based concept generation. The use of the methods in a straightforward manner highlights that many methods can be "plugged in" and that the results are attributed to the model's process of combination and verification. Our model combines the above methods since each method presents a unique advantage

- internal perspective of the web service by the TF/IDF,
- external perspective of the web service by the Web Context
- Extraction, and a comparison to a free text description

Figure 3.1 describes the ontological bootstrapping process for web services.

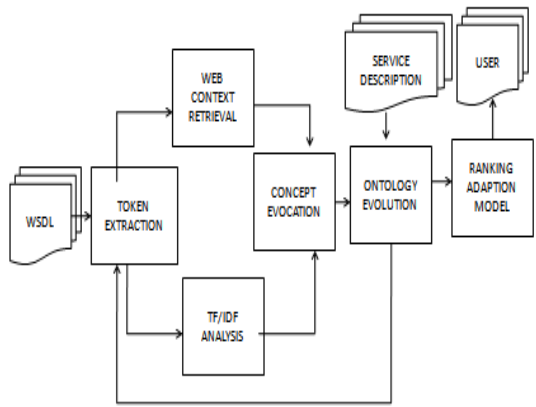

Figure 3.1 Web Service Ontology Bootstrapping Process

## 4. CONTEXT ANALYSIS IN ONTOLOGICAL BOOTSTRAPPING

Context analysis involves the following processes
- Token extraction
- TF/IDF analysis
- Web Context extraction and

Token extraction process extracts tokens representing relevant information from a WSDL document. It extracts all the name labels, parses the tokens, and performs initial filtering. WSDL tokens are composed of a sequence of words with each first letter of the words are capitalized. Therefore, the descriptors are divided into separate tokens. The sequence of words is expanded using the capital letter of each word. The tokens are filtered using a list of stop words, removing words with no substantive semantics.

In TF/IDF analysis process, the most common terms appearing in each web service document and appearing less frequently in other documents are analyzed. TF/IDF weight is a statistical measure used to evaluate how important a word is to a document in a collection. If a term or word appears lots in a document but also appears lots in the collection as a whole it will get a lower score. The method is applied here to the WSDL descriptors. [2]

The web context extraction process uses the sets of tokens as a query to a search engine, clusters the results according to textual descriptors, and classifies which set of descriptors identifies the context of the web service. The context extraction process is defined as tokens extracted from the web service WSDL descriptor. Each set of tokens is then sent to a web search engine and a set of descriptors is extracted by grouping the web pages search results for each token set.

Figure 4.1 describes the context analysis for WSDL documents.

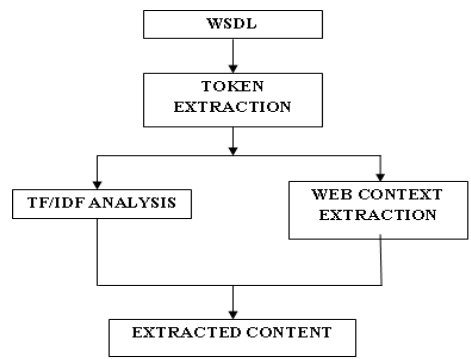

Figure 4.1 Context analysis for WSDL Document

## 5. STEPS INVOLVED IN ONTOLOGICAL BOOTSTRAPPING PROCESS

The ontology bootstrapping involves the process
- Concept evocation
- Ontology evolution
- Ranking adaption

The Concept evocation process identifies a possible concept definition. The concept evocation is performed based on context intersection. An ontology concept is defined by the descriptors that appear in the intersection of both the web context results and the TF/IDF results.

The ontology evolution expands the ontology as required according to the newly identified concepts and modifies the relations between them. The external web service textual descriptor serves as a moderator if there is a conflict between the current ontology and a new concept. New concepts can be checked against the free text descriptors to verify the correct interpretation of the concept. After the ontology evolution, the whole process continues to the next WSDL with the evolved ontology concepts and relations. [2]

Ranking adaptation is comparatively more challenging. It desires to adapt the model which is used to predict the rankings for a collection of web service documents. Though the web service documents are normally labeled with several relevance levels, which seem to be handled by a multiclass classification or regression, this process categorizes the web service queries based upon most related searched web service. Most visited web service is categorized based upon priority [1]

Figure 5.1 illustrates the process of ontology generation and rank assignment of web services according to user request.

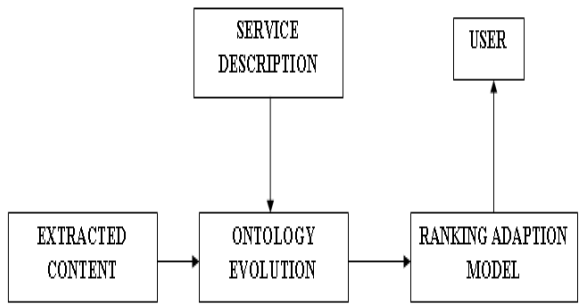

Figure 5.1 Ontology Generations and Rank Assignment for Web Services

## 6. CONCLUSION

This paper proposes an approach for bootstrapping an ontology based on web service descriptions. The approach analyzes web services from multiple perspectives and integrates the results. Web services usually consist of both WSDL and free text descriptors. This allows bootstrapping the ontology based on WSDL and verifying the process based on the web service free text descriptor.

Ontology bootstrapping is used to focus on unlimited domain so that any type of web service can be created, maintained and added in the existing ontology. This approach facilitates ontology genration process which allows the automatic building of an ontology that can assist in expanding, classifying and retrieving relevant services without any guidance as a contrary to the existing methods.

The ontology bootstrapping process in our model is performed automatically, enabling a constant update of the ontology for every new web service. Using ranking adaption model, most frequently visited web service can be easily identified. Thus, the automatic construction, enrichment and adaptation of ontologies, is highly preferred.